\documentclass[twocolumn]{ieeetran}    

\usepackage{graphicx,here,bm,multirow}
\usepackage{amsmath}
\usepackage{amssymb}
\usepackage{times}
\usepackage{amsfonts}
\usepackage{enumerate}
\usepackage{longtable}
\usepackage{graphicx}
\usepackage{theorem}
\usepackage{multirow}
\usepackage{mdframed}
\usepackage{booktabs}
\usepackage{float}    
\usepackage{array} 
\usepackage[round,sort,semicolon]{natbib}

\newtheorem{remark}{Remark} [section]

\usepackage{tikz}
\usetikzlibrary{arrows,shapes,calc}
\usepackage{pgfplots}
\usetikzlibrary{positioning}
\usetikzlibrary{pgfplots.groupplots}

\usepackage{tikz}
\usetikzlibrary{arrows,shapes,calc}
\usepackage{pgfplots}
\usetikzlibrary{positioning}

\usepackage[utf8]{inputenc}
\usepackage{pgfplots}

\title{On the use of supervised clustering in stochastic NMPC design}

\author{Mazen Alamir \thanks{The material in this paper was not presented at any conference. M. Alamir is with Univ. Grenoble Alpes, CNRS, Grenoble INP*, GIPSA-lab, 38000 Grenoble, France. Email: mazen.alamir@grenoble-inp.fr  (http://www.mazenalamir.fr).}
}

\begin{document}




\maketitle

\begin{abstract}                          
In this paper, a supervised clustering based-heuristic is proposed for the real-time implementation of approximate solutions to stochastic nonlinear model predictive control frameworks. The key idea is to update on-line a low cardinality set of uncertainty vectors to be used in the expression of the stochastic cost and constraints. These vectors are the centers of uncertainty clusters that are built using the optimal control sequences, cost and constraints indicators  as supervision labels. The use of a moving clustering data buffer which accumulates recent past computations enables to reduce the computational burden per sampling period while making available at each period a relevant amount of samples for the clustering task. A relevant example is given to illustrate the contribution and the associated algorithms. 
\end{abstract}

\section{Introduction}
Stochastic Nonlinear Model Predictive Control (SNMPC) is without doubt one of the major challenges facing the NMPC community for the years to come. This can be viewed as the third {\em key step} to achieve. Indeed, after the 90s where the provable stability was the main paradigm \citep{Mayne2000}, the last ten years or so were dedicated to making available reliable and easy to use NMPC solvers for nominal deterministic settings \citep{Andersson2018}. The success of these two steps helped propelling MPC-based solutions {\em out-of-labs} towards the real-life paradigm where the keywords are {\em risk}, {\em uncertainties} and {\em probability}. 

After some early attempts involving Robust NMPC \citep{Magni2007} which rapidly appeared to be over stringent, it quickly becomes obvious that the natural way to address the new paradigm is to replace all the MPC ingredients (cost, constraints) by their {\bf expected} counterparts in the formulation of the open-loop optimization problem. Stochastic NMPC was born for which excellent recent unifying reviews can be found in \citep{Mesbah2016, MAYNE2016184,Mesbah:2018}. 

Unfortunately, the apparently intuitive and simple shift in paradigm consisting in doing the {\em business as usual} on the expected quantities, comes with heavy consequences in terms of computational burden. Indeed, computing the expectation of a nonlinear function of several variables for each candidate control sequence is obviously an impossible task. Only approximations can be attempted, each coming with its own merits and drawbacks. 

The first idealistic option is to use the Stochastic Dynamic Programming (SDP) framework which is based on the well known Bellman's principle of optimality in which the conditional probability plays the role of extended state \citep{Mesbah:2018}. Unfortunately, solving the SDP leads to algorithms that scale exponentially in the dimension of the state. Nevertheless, for small sized problems, nice and elegant solutions can be derived \citep{Rigaut:2018} that might even address realistic real-life problems.

A second option is to derive on-line a structured approximation (Gaussian Processes or chaos polynomials for instance) of the probability density function (pdf) at the current state and then to use the resulting approximation in evaluating the expectation of relevant quantities \citep{Bradford:2018, NAGY2007229}. Note however that this has to be done for all possible candidate control sequences in each iteration of the NLP solver. This obviously restricts the field of application of this approach to small-sized and rather slow systems if any. 

The third and probably more pragmatic option is to use scenarios-based averaging in order to approximate the expectations (or optionally higher order moments) involved in the problem formulation \citep{SCHILDBACH20143009}. In this case, a high number (say $K$) of samples of the random quantities is drawn and the resulting constraints and state equations are concatenated while sharing the same control. A common {\em optimal} control sequence is then searched for using standard nominal solvers. 

This last approach may lead to a very high dimensional problem that is not intuitively prone to a parallel computing or distribution over the system life-time. This is especially true when the underlying (deterministic problem) is solved using efficient multiple-shooting algorithms \citep{Bock:2000} since the dimension of the extended state is proportional to the number of samples $K$ being involved. The latter can be quite high in order to get a decent level of certification \citep{alamo2009}. Moreover, the need to introduce variance related terms in the formulation to better address the chance constraint certification \citep{Mesbah:2014} makes things even worse as double summation on the set of scenarios has to be performed leading to a $K^2$-rated complexity.

It is worth underlying that even when putting aside the computational challenges associated to SNMPC, one has to keep in mind that all these methods assume that the statistical description of the uncertainty vector is available (to draw relevant samples) and that the problem lies in the way to propagate it depending on the control actions. This knowledge is never available and can only be presumed. This should achieve convincing us that we need to accept a painful transition from a proof-related certain paradigm to a realm of heuristics which can only be evaluated once implemented and its results diagnosed on real-life problems. Consequently, the implementability/Scalability issues become crucial and key properties of any solution framework to SNMPC. 

The present paper addresses the scenario-based SNMPC framework under this last point of view, namely, that of implementable and scalable heuristics. 

\begin{figure}
\begin{center}
\begin{tikzpicture}
\node[draw, thick,rounded corners,inner xsep=2mm, inner ysep=2mm,fill=black!5] at(0,0)(SNMPC){\begin{minipage}{0.22\textwidth}
\begin{center}
{\bf Scenario-Based SNMPC} \\
{\small $n_{cl}$ disturbance representatives $\{(\bm w^{(i)},p^{(i)},\sigma_J^{(i)},\sigma_g^{(i)})\}_{i=1}^{n_{cl}}$}
\end{center}
\end{minipage}};
\node[draw,rounded corners,inner xsep=2mm, inner ysep=5.5mm] at(3.5,0)(SYST){System};
\node[draw,rounded corners,inner xsep=4mm, inner ysep=3mm] at(0,2)(CLS){{\bf Clustering} {\small ($n_{cl}$ classes)}};
\node[draw,rounded corners,inner xsep=2mm, inner ysep=3mm] at(0,4.5)(BUFF){\begin{minipage}{0.25\textwidth}
\begin{center}
{\bf FIFO-Buffer} \\ 
{\footnotesize ($n_b=q\cdot N_n$ data): optimal solutions for samples in the buffer $\mathcal D_b:=\{\bm w_b^{(s)},\bm u_*^{(s)}, J_*^{(s)}, g_*^{(s)}\}_{s=1}^{n_b}$}
\end{center}
\end{minipage}};
\node[draw,rounded corners,inner xsep=2mm, inner ysep=3mm] at(0,7.4)(NMPC){\begin{minipage}{0.27\textwidth}
\footnotesize New uncertainties $\{\bm w^{[j]}\}_{j=1}^{N_n}$ \\
solve in {\bf Parallel} NMPC$(w^{[j]})\vert x$\\
stack the result to get \\
$\mathcal D_n:=\{(\bm w^{[j]},\bm u_*^{[j]},J_*^{[j]},g_*^{[j]})\}_{j=1}^{N_n}$
\end{minipage}};

\draw[->,>=stealth] (SNMPC) -- node[midway, above]{$u$}(SYST);
\draw[->,>=stealth] (CLS.198) -- node[midway, right]{\footnotesize $\mathcal D:=\{\bm w^{(i)},p^{(i)},\sigma_J^{(i)},\sigma_g^{(i)}\}_{i=1}^{n_{cl}}$} ++(0,-0.8);
\draw[->,>=stealth] (BUFF.217) -- node[midway, right]{\footnotesize $\mathcal D_b\leftarrow {\rm FIFO}(\mathcal D_b,\mathcal D_n)$} ++(0,-0.95);
\draw[->,>=stealth] (NMPC) -- node[midway, right]{\footnotesize $\mathcal D_n$} (BUFF.north);
\draw[->,>=stealth] (SYST) |- node[midway, right]{$x$} (NMPC.east);
\end{tikzpicture}
\end{center}
\caption{Schematic view of the proposed SNMPC framework.} \label{schematic}
\end{figure}
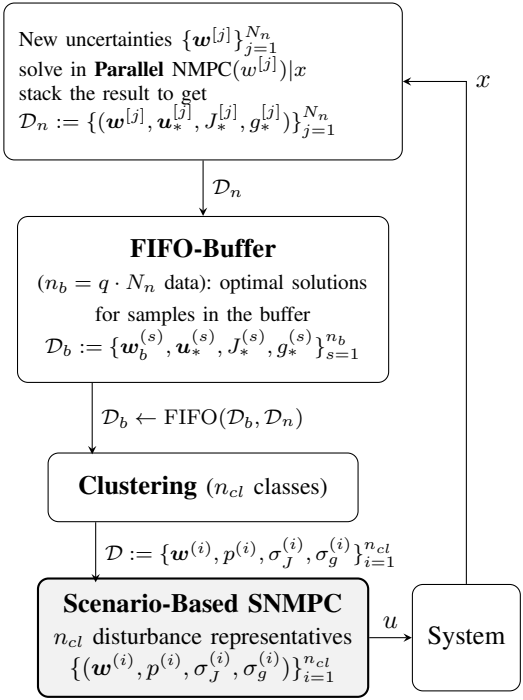

An overview of the framework proposed in the present contribution is sketched in Figure \ref{schematic}. In this Figure, $x$, $u$, $w$, $J$ and $g$ refer to state, control, uncertainty, cost function and constraint respectively. The basic block (at the bottom of Figure \ref{schematic}) where SNMPC is performed is the grayed box that delivers the action to be applied to the controlled system, namely, the first action in the scenario-based optimal sequence. 

The key difference with the standard implementation is that the SNMPC is formulated using only a few number ($n_{cl}$) of regularly updated disturbance samples. More precisely, $n_{cl}$ is the number of classes used in a clustering step. This clustering box delivers to the SNMPC box a regularly updated set containing the centers of clusters together with their population weights ($p^{(i)}$) and dispersion indicators $(\sigma_J^{(i)}, \sigma_g^{(i)})$ in the data set $\mathcal D_b$ used to achieve the clustering task. This data set $\mathcal D_b$ is accumulated in a First In First Out (FIFP) buffer. The latter receives at each updating step a new block of data $\mathcal D_n$ which is delivered by the top block. This data bloc $\mathcal D_n$ contains a set of $N_n$ nominal solutions $\bm u_*^{[j]}$ of a standard NMPC with presumably known newly sampled disturbance vectors $w^{[j]}$ together with the corresponding optimal costs and constraints indicators $J_*^{[j]}$, $g_*^{[j]}$, $j=1,\dots,N_n$. As the $N_n$ optimization problems are totally decoupled, the computation performed in this top bloc can be done in fully parallel way.  

The rational behind this framework lies in the intuition that very often, while the space of possible uncertainty realizations might be very rich (including uniform distributions in high dimensional hypercubes), the set of corresponding optimal ingredients (control sequences, optimal cost, constraints indicators) might accept a low cardinality set of meaningfully distinct clusters. Moreover, the loss of information that results from using only the centers of clusters in the formulation can be partially mitigated by using the statistical information $(\sigma_J^{(i)},\sigma_g^{(i)})$, $i=1,\dots,n_{cl}$ regarding the dispersion of cost and constraints indicator within each cluster. This information is transmitted from the clustering layer as indicated in Figure \ref{schematic}. Sections \ref{secBuffer} and \ref{secclustering} give more detailed description of the above two steps. 

The aim of this paper is to give a rigorous presentation of this framework and to propose a complete implementation on a relevant example in order to assess the performance and implementability of the framework. 

This paper is organized as follows: Section \ref{secNota} gives some definitions and notation used in the sequel. The Proposed framework is explained in Section \ref{secproposed} by successively explaining the different boxes depicted in Figure \ref{schematic}. An illustrative example is given in Section \ref{secexample}. Finally, Section \ref{secConc} concludes the paper and gives some hints for further investigation. 
\section{Definitions and notation} \label{secNota}
We consider nonlinear dynamic systems given by
\begin{equation}
x^+=f(x,u,w) \label{syst}
\end{equation}
where $x\in \mathbb{R}^{n}$, $u\in \mathbb U\subset \mathbb{R}^{m}$ and $w\in \mathbb{R}^{r}$ stand respectively for the vectors of state, control and uncertainty. It is assumed here for simplicity that the whole state vector is measured while the uncertainty is not. Moreover, it is also assumed that the size of the uncertainty vector and the level of excitation are such that the uncertainty estimation through dedicated observer is not a reasonable  option. 

Consider that a couple of cost/constraints functions can be defined at any current state $x$ by\footnote{Boldfaced notation $\bm x$, $\bm u$ and $\bm w$ are used to denote variables profiles over some prediction horizon.} $J^{(x)}(\bm u,\bm w)\in \mathbb{R}_+$ and $g^{(x)}(\bm u, \bm w)\le 0\in \mathbb{R}$ that expresses respectively a cost function to be minimized (in some sense) and a constraints violation indicator to be limited (in some sense) over some finite prediction horizon of length $N$ and starting from the initial state $x$. When the state is implicitly known (or fixed during some argumentation), the short notation $J(\bm u,\bm w)$ and $g(\bm u,\bm w)$ can be used. 

\begin{remark}
Note that $g$ is a scalar map that might encompass a set of constraints to be enforced through dedicated maps (such as $\max\{0,\cdot\}^2$ for instance). The treatment of this function can be vectorized for computational efficiency (including by using of a vector of slack variables in softening the constraints rather than the scalar used in the sequel). We keep nevertheless this scalar notation for the sake of simplicity of exposition of the main ideas. In the simulation however, vectorized implementation is used. 
\end{remark}
The ideal stochastic NMPC formulation that is approximated in the present paper takes the following form:
\begin{align}
&\min_{(\bm u,\mu)} \quad \mathbb E(J^{(x)}(\bm u,\cdot))+\left[\frac{1-\epsilon_J}{\epsilon_J}\right]\mathbb S(J^{(x)}(\bm u,\cdot))+\rho\mu\label{P1} \\
& \mbox{\rm under}\ \mathbb E(g^{(x)}(\bm u,\cdot))+\left[\frac{1-\epsilon_g}{\epsilon_g}\right]\mathbb S(g^{(x)}(\bm u,\cdot))\le \mu \ge 0 \label{P2}
\end{align}
which can be understood by means of the following comments:\\ \ \\ 
$\checkmark$ $\mathbb E$ and $\mathbb S$ denote respectively the expectation and the standard deviation of their arguments over the presumably known statistics on the uncertainty vector $\bm w$.\\ \ \\ 
$\checkmark$ According to \citep{Mesbah:2014}, when $\mu=0$  the satisfaction of (\ref{P2}) implies that the probability of satisfaction of the original constraint $g^{(x)}(\bm u,\bm w)\le 0$ is greater than $1-\epsilon_g$ and this, regardless of the specific statistics of the uncertain variables. Using $\mu$ with a high penalty $\rho$ implements a soft version of this formulation. 
\\ \ \\ 
$\checkmark$ Similarly, the cost function that is minimized in (\ref{P1}) when using $\rho=0$, is precisely the bound below which it can be certified, with a probability greater than $1-\epsilon_J$, that the expectation of the cost lies.
\\ \ \\ 
The difficulty in implementing a solution to the formulation (\ref{P1})-(\ref{P2}) lies in the cost of approximating the expectation and standard deviation involved. The commonly used approach replaces the expectation by an averaging sum over a high number of uncertainties samples which can be quite heavy to compute as mentioned in the introduction. In the following section, the proposed approximating method to the formulation (\ref{P1})-(\ref{P2}) is described. 
\section{The proposed framework} \label{secproposed}
In this section, the different tasks involved in the framework depicted in Figure \ref{schematic} are successively detailed. 
\subsection{Solving a set of deterministic problems: Construction of a new data set $\mathcal D_n$} \label{sec-Dn}
This task consists in drawing a new set of $N_n$ values $\bm w^{[j]}$ of the uncertainty profile and to solve, knowing these values, the corresponding individual deterministic constrained optimization problem given by:
\begin{align}
\bm u_*^{[j]}\leftarrow \min_{\bm u}\quad J^{(x)}(\bm u,\bm w^{[j]})+\rho\mu^2\ \vert\ g^{(x)}(\bm u,\bm w^{[j]})\le \mu \label{nominalPb}
\end{align}
the resulting individual optimal cost and constraints are denoted by $J_*^{[j]}$ and $g_*^{[j]}$ respectively. This enables the following data set to be defined:
\begin{equation}
\mathcal D_n:=\{(\bm w^{[j]},\bm u_*^{[j]},J_*^{[j]},g_*^{[j]})\}_{j=1}^{N_n} \label{defdeDn}
\end{equation}
Note that solving these individual problems while knowing the values of the disturbance profiles enables to reveal a {\em population of control sequences} that would be optimal should the disturbance profiles that originates them occurs. The relevance of this computation is to use the resulting data in a disturbance-profiles clustering step. This is because: \\ \ \\ 
\begin{minipage}{0.05\textwidth}
\ 
\end{minipage}
\begin{minipage}{0.02\textwidth}
\color{gray}\rule{1mm}{25mm}
\end{minipage}
\begin{minipage}{0.35\textwidth}
The disturbance profiles that correspond to {\em similar} optimal control sequences, optimal cost and constraint values, should be declared to lie in the same cluster of disturbance profiles even if they strongly differ as a high dimensional vectors. 
\end{minipage}
\\ \ \\ \ \\
Because the clustering is based on the labels constituted by the triplet $(\bm u_*^{[j]}, J_*^{[j]}, g_*^{[j]})$, the clustering is qualified hereafter as a supervised clustering. 

Note that this step is totally parallelizable as the individual deterministic problems are totally decoupled. Nevertheless, the number $N_n$ of samples can be moderate since a buffer is created and updated by such data at each sampling period as explained and justified in the next section. 

Since the dataset $\mathcal D_n$ is related to a current state $x_k$ at instant $k$, it is denoted by $\mathcal D_n(k)$ when the reference to the sampling instant $k$ is needed. 
\subsection{Updating the clustering buffer: Creating and updating the dataset $\mathcal D_b$} \label{secBuffer}
This is a simple FIFO data storage task in which the successive datasets $\mathcal D_n$ of the form (\ref{defdeDn}) are stacked for use in the clustering task. 

As the new datasets $\mathcal D_n(k)$ are added at each sampling period $k$, the size $n_b:=qN_n$ of the clustering buffer ($q$ is the number of successive datasets $\mathcal D_n(k)$ to be included) depends on the bandwidth of the system. This is because integrating all the datasets $\mathcal D_n(k)$  in a single clustering dataset (called $\mathcal D_b$ in Figure \ref{schematic}) ignores the fact that each of these datasets is related to a different state that defines the underlying optimization problem (\ref{nominalPb}). The underlying assumption is that the evolution of the state during the $q$ successive sampling periods can be viewed as sufficiently small for the clustering dataset $\mathcal D_b$ to remain relevant. 

To summarize, at each sampling period $k>q$, the clustering data set is given by:
\begin{equation}
\mathcal D_b(k) := \{\mathcal D_n(k-1),\mathcal D_n(k-1),\dots,\mathcal D_n(k-q)\} \label{defdeDb}
\end{equation}
where $\mathcal D_n(k-j)$ is the dataset containing the solutions of the $N_n$ nominal problems defined by (\ref{nominalPb}) with the state $x_{k-j}$. For smaller initial values of $k$, the buffer contains only the available $k-1$ datasets $\mathcal D_n(k-1),\dots,\mathcal D(0)$.

\begin{remark}
The choice of the size $q$ of the clustering set $\mathcal D_b$ is obviously the object of a recurrent type of dilemmas commonly encountered in real-time MPC. This dilemma holds  between the quality of the solution of a problem (better if the size of the cluster is large) and the very relevance of the problem itself (weak if too long computation time is used before updating the action accordingly). In nominal deterministic MPC, the parameter to be tuned is the number of iterations of the underlying optimization algorithm \citep{Alamir:2017}. 
\end{remark}
Having the clustering data $\mathcal D_b$, the next section explains the supervised clustering task that leads to the selection of the $n_{cl}$ clusters whose centers form the database $\mathcal D$ feeding the SNMPC formulation (Figure \ref{schematic}). 
\subsection{Clustering the uncertainty set: Creating and updating the low cardinality dataset $\mathcal D$} \label{secclustering}
Clustering is a key branch of Data Mining whose objective is to split a set of data into {\em subsets} such that inside each subset, the data are similar in some sense (according to some distance). Obviously, the clustering topic is vast and it is outside the scope of the present contribution to give a survey of available clustering techniques. Readers can consult \citep{Xu2015} for a comprehensive and recent survey. 

Fortunately enough, when it comes to use clustering (or more generally many Machine Learning) algorithms as parts of a wider solution framework (as it is the case in the present contribution), free publicly available implementations of clustering task can be used such as the well-known scikit-learn library \citep{scikit-learn}.

A clustering map $C$ takes as arguments: 
\begin{itemize}
\item  a disctere set $\mathcal V:=\{v^{(i)}\}_{i=1}^{n_b}$ to be split into clusters
\item an integer $n_{cl}$ representing the number of clusters which one wishes $\mathcal V$ to be split into,
\end{itemize}
and delivers as output a $n_b$-dimensional vector of {\em labels} $\mathcal I\in \{1,\dots,n_{cl}\}^{n_b}$ that associates to each member $v_i$ of $\mathcal V$ its associated cluster. This is shortly written as follows:
\begin{equation}
\mathcal I=C(\mathcal V,n_{cl})\in \{1,\dots,n_{cl}\}^{n_b}
\end{equation}
Recall that our objective is to perform a clustering of the set of disturbance vectors $\mathcal W:=\{w^{(s)}_b\}_{s=1}^{n_b}$ contained in the dataset $\mathcal D_b$ (see Figure \ref{schematic}).

Clustering algorithms (K-Means, Mean-shift, DBSCAN, to cite but few algorithms in the scikit-learn library) generally perform a {\em unsupervised learning} in the sense that they consider only internal relationships and distances between the elements of the set $\mathcal V$ to split and this regardless of any exogenous information\footnote{Called {\em labels} in the Machine Learning language.} about these elements.  

Following the discussion of section \ref{secproposed}, we seek a clustering that considers as {\em similar} those disturbance vectors that correspond to {\em similar} triplets of control profiles, cost and constraint indicators. This is the reason why the set $\mathcal V$ that is used hereafter is given by:
\begin{equation}
\mathcal V = \{\bm (u_*^{(s)},J_*^{(s)},g_*^{(s)})\}_{s=1}^{n_b} \label{defdecalV}
\end{equation}
That is why we refer to the proposed clustering approach as a {\em supervised clustering} as the set of class labels $\mathcal I$ that will be used to split the uncertainty vectors set is derived using the {\em exogenous} information contained in the set $\mathcal V$ given by (\ref{defdecalV}), namely:
\begin{equation}
\mathcal I:= C\Bigl(\{\bm (u_*^{(s)},J_*^{(s)},g_*^{(s)})\}_{s=1}^{n_b}, n_{cl}\Bigr)
\end{equation}
Once this clustering is achieved, the centers of the $n_{cl}$ resulting clusters are given as a by-side product of the clustering task:
\begin{equation}
w^{(i)}:={\rm Mean}\left(w_b^{(s)}, s\in \{1,\dots,n_b\}\ \vert\  \mathcal I_s=i\right)
\end{equation}
Beside these centers, the weights of the different clusters can be associated to the relative size of their populations, namely:
\begin{equation}
p^{(i)}:=\dfrac{1}{n_b}{\rm card}\left\{s\in \{1,\dots,n_b\}\ \vert \ \mathcal I_s=i\right\}
\end{equation}
Finally, evaluations of the dispersions of the cost function and the constraints inside each cluster can also be cheaply obtained using the two variances defined by:
\begin{align}
\sigma_J^{(i)}&:={\rm Var}\left(J_*^{(s)}, s\in \{1,\dots,n_b\}\ \vert\  \mathcal I_s=i\right) \label{varJ}\\
\sigma_g^{(i)}&:={\rm Var}\left(g_*^{(s)}, s\in \{1,\dots,n_b\}\ \vert\  \mathcal I_s=i\right) \label{varJ}
\end{align}
This ends the definition of the dataset $\mathcal D$ (see Figure \ref{schematic}) that is used in the formulation of the SNMPC which is described in the following section. 
\subsection{Formulation of the stochastic NMPC} \label{secSNMPC}
In this section, approximate expressions for (\ref{P1}) and (\ref{P2}) are given using the ingredients contained in the dataset $\mathcal D$ which is updated at each sampling period using the steps explained in the previous sections. This is done by averaging over the set of centers $w^{(i)}$, $i=1,\dots,n_{cl}$ of the clusters created above while accommodating for the dispersion inside the clusters. More precisely, the following optimization problem is considered
\begin{align}
&\min_{\bm u, \mu} \sum_{i=1}^{n_{cl}}p^{(i)}\left[J^{(x)}(\bm u,w^{(i)})+\dfrac{1-\epsilon_J}{\epsilon_J}\sqrt{\sigma_J^{(i)}}+\rho\mu\right] \label{defdeEaJ}\\
& \mbox{\rm under the following constraints defined for $\forall i\in \{1,\dots,n_{cl}\}$} \nonumber \\
&g^{(x)}(\bm u,w^{(i)})+\dfrac{1-\epsilon_g}{\epsilon_g}\sqrt{\sigma_g^{(i)}}\le \mu \ge 0\label{defdeEag}
\end{align}
where a $p^{(i)}$-based weighted sums in which the predicted values at the center of the clusters are augmented by the terms that depend on the estimated variances $\sigma^{(i)}_J$ and $\sigma_g^{(i)}$ included in the dataset $\mathcal D$ as described above. 
\begin{remark}
Note that thanks to the low number of clusters $n_{cl}$, one can afford to enforce the personalized approximated expression of (\ref{P2}) on each cluster individually rather than taking the global statistics over all the clusters. 
\end{remark}

\section{Illustrative Example: Combined therapy of cancer} \label{secexample}
\subsection{System equations, objective and constraints}
As an illustrative example, let us consider the problem of drug dosing during a combined chemotherapy/immunotherapy of cancer \citep{KASSARA2011135,ALAMIR201559}. The dynamic model four states, two control inputs and 13 uncertain parameters. More precisely, the state components are defined as follows:
\begin{tabbing}
\hskip 1cm \= \hskip 6cm \kill 
$x_1$ \> tumor cell population \\
$x_2$ \> circulating lymphocytes population\\
$x_3$ \> chemotherapy drug concentration \\
$x_4$ \> effector immune cell population \\
$u_1$ \> rate of introduction of immune cells \\
$u_2$ \> rate of introduction of chemotherapy 
\end{tabbing}
and the dynamics is given by:
\begin{align}
\dot x_1&=ax_1(1-bx_1)-c_1x_4x_1-k_3x_3x_1 \label{model1} \\
\dot x_2&=-\delta x_2-k_2x_3x_2+s_2 \label{model2} \\
\dot x_3&=-\gamma_0 x_3+u_2 \label{model3} \\
\dot x_4&=g\dfrac{x_1}{h+x_1}x_4-rx_4-p_0x_4x_1-
k_1x_4 x_3+s_1u_1 \label{model4}
\end{align}  
The description of the relevance of each term and the coefficient can be examined in \citep{ALAMIR201559} although one can easily guess from the definition of the state components. \\ \ \\ 
Using the notation above, the uncertainty vector $w$ gather all the uncertain parameters involved in the model (\ref{model1})-(\ref{model4}), namely:
\begin{equation}
w:= \begin{bmatrix}
a,b,c_1,k_3,\delta,k_2,s_2,\gamma_0,g,r,p_0,k_1,s_1
\end{bmatrix}\in \mathbb{R}^{13}_+ \label{defw}
\end{equation}
that are supposed here to be constant but unknown. Note also that a reconstruction of all these parameters from patient measurement during the treatment is obviously out of question. Table \ref{tab2} gives the nominal values of the parameters involved in the dynamics. Note that because of the excursion of these parameters and the related states, a normalized version of the dynamics (\ref{model1})-(\ref{model4}) is used by using the following vector of reference state:
\begin{equation}
\bar x := \begin{bmatrix}
10^{9}, 10^9, 1, 10^9
\end{bmatrix} 
\end{equation}
\begin{table*}
\begin{center}
\begin{tabular}{llllll} \toprule
    param & value & param & value  & param & value\\   \midrule  
    $a$ & $0.25\ day^{-1}$ & $b$ & $1.02\times 10^{-14}\ cell^{-1}$ & $c_1$ & $4.41\times 10^{-10}\ (cell\cdot day)^{-1}$\\
 $g$ & $1.5\times 10^{-2}\ day^{-1}$ & $h$& $2.02\times 10^1\ cell^2$ & $k_2,k_3$ & $6\times 10^{-1}\ day^{-1}$ \\ $k_1$ & $8\times 10^{-1}\ day^{-1}$ & $p_0$ & $2\times 10^{-11}\ (cell\cdot day)^{-1}$ &
$s_1$ & $1.2\times 10^7\ cell\cdot day^{-1}$ \\ $s_2$ & $7.5\times 10^6\ cell\cdot day^{-1}$ & $\delta$ & $1.2\times 10^{-2}\ day^{-1}$ &
$\gamma$ & $9\times 10^{-1}\ day^{-1}$\\
$r$ & $4.0\times 10^{-2}\ cell\cdot day^{-1}$\\
\bottomrule
\end{tabular}
\end{center} 
\ \\
\caption{Nominal values of the parameters.}\label{tab2} 
\end{table*}
As it is typically the case in cancer treatment, the control objective is to reduce the tumor cells population $x_1$ at the end of the treatment while ensuring that the health of the patient (represented in the above model by the circulating lymphocytes population size $x_2$) remain greater than some a priori fixed lower bound $x_2^{min}$.\\ \ \\ 
Consequently, the following cost function is used at each state $x$ in the MPC design:
\begin{equation}
J^{(x)}(\bm u,w) := \rho_f x_1(N) + \sum_{i=1}^N x_1(i\vert \bm u, x, w)+\rho_u \vert u(i)\vert 
\end{equation}
together with the following constraint to be enforced on the predicted trajectory:
\begin{equation}
g^{(x)}(\bm u,w):= \min_{i\in \{1,\dots,N\}}\left[x_2(i\vert \bm u, x, w)\right]\ge x_2^{min}
\end{equation}
The control input is saturated according t $u\in [0,5]\times [0,1]$. 
\subsection{The stochastic MPC controller settings}
In all the forthcoming simulations, the sampling period $\tau=0.2$ (Days) is used. When Stochastic MPC is used, the number of clusters is taken equal to $n_{cl}=3$. The number $N_n=25$ of new samples is generated at each sampling period (see Figure \ref{schematic}). The size of the FIFO buffer is taken equal to $n_b=4\times 25=100$ ($q=4$). The parameters $\epsilon_J$ and $\epsilon_g$ used to account for the variance in the definition of the cost function and the constraints are taken equal to $\epsilon_J=\epsilon_g=0.1$ (leading to $90\%$ of confidence rate). The weighting coefficients $\rho_f=1000$, $\rho_u=1$ and $\rho=10$ are used. The clustering is performed using the KMeans module of the scikitlearn python library \citep{scikit-learn}.
\\ \ \\ 
The stochastic MPC is compared to the nominal MPC which uses the nominal values of the parameters as given in Table \ref{tab2}. As for the stochastic MPC, the random values of these parameters are obtained according to:
\begin{equation}
w_i=(1+\nu_i)\bar w_i\quad \mbox{\rm where $\nu_i\in \mathcal N(0,\sigma)$}
\end{equation}
where a variance $\sigma=0.2$ is used leading to samples than might have a discrepancy that might be as high as 45-80\% of the nominal values. $100$ simulations are performed using either stochastic or nominal MPC and statistical indicators are compared. Note that the cloud of disturbances used in these $100$ simulations are {\em fired} independently of those fired to feed the FIFO buffer of the stochastic MPC. All the simulations use the normalized initial state $x_0=(1.0,0.15,0,1)$ and all the simulations last 40 Days. The prediction horizon length is taken equal to $N=10$ (2 Days) and five steps of the optimal control sequence is applied before a new optimal sequence is computed. This leads to an updating control period of $1$ Day. The problem encoding and the optimization are performed using multiple-shooting formulation (with hot starting of the initial guess at each sampling period) free software CasADi \citep{Andersson2018} (python version) on a MacBookPro 2.9 GHz Intel Core i7.
\subsection{Results and discussion}
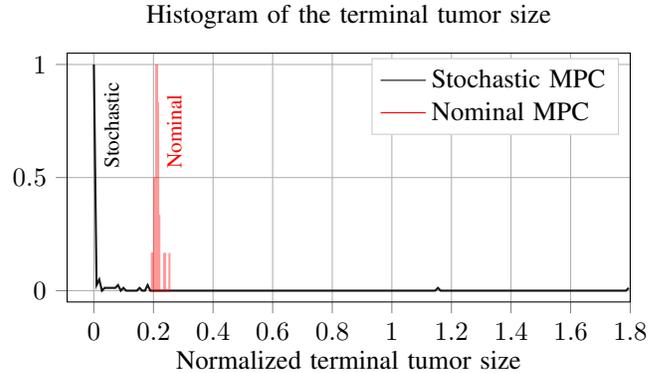
\begin{figure}[H]
\begin{center}
\begin{tikzpicture}

\definecolor{color0}{rgb}{0.12156862745098,0.466666666666667,0.705882352941177}
\definecolor{color1}{rgb}{1,0.498039215686275,0.0549019607843137}

\begin{axis}[
title={Histogram of the terminal tumor size},
xlabel={Normalized terminal tumor size},
xmin=-0.0897496577590674, xmax=1.8,
ymin=-0.05, ymax=1.05,
tick align=outside,
tick pos=left,
xmajorgrids,
width=0.5\textwidth,
height=0.2\textheight,
x grid style={white!69.01960784313725!black},
ymajorgrids,
y grid style={white!69.01960784313725!black},
legend style={draw=white!80.0!black},
legend entries={{Stochastic MPC},{Nominal MPC}},
legend cell align={left}
]
\addlegendimage{no markers, black}
\addlegendimage{no markers, red}
\addplot [thick, black, opacity=0.9]
table {%
8.69111140678573e-14 1
0.00902006610653458 0.0246913580246914
0.0180401322129822 0.0493827160493827
0.0270601983194299 0
0.0360802644258776 0.0123456790123457
0.0451003305323252 0.0123456790123457
0.0541203966387729 0.0123456790123457
0.0631404627452206 0.0123456790123457
0.0721605288516682 0.0123456790123457
0.0811805949581159 0.0246913580246914
0.0902006610645636 0
0.0992207271710112 0.0123456790123457
0.108240793277459 0
0.117260859383907 0
0.126280925490354 0
0.135300991596802 0
0.14432105770325 0
0.153341123809697 0.0123456790123457
0.162361189916145 0
0.171381256022593 0
0.18040132212904 0.0246913580246913
0.189421388235488 0
0.198441454341936 0
0.207461520448383 0
0.216481586554831 0
0.225501652661279 0
0.234521718767726 0
0.243541784874174 0
0.252561850980622 0
0.261581917087069 0
0.270601983193517 0
0.279622049299965 0
0.288642115406412 0
0.29766218151286 0
0.306682247619308 0
0.315702313725755 0
0.324722379832203 0
0.333742445938651 0
0.342762512045098 0
0.351782578151546 0
0.360802644257994 0
0.369822710364441 0
0.378842776470889 0
0.387862842577337 0
0.396882908683784 0
0.405902974790232 0
0.41492304089668 0
0.423943107003127 0
0.432963173109575 0
0.441983239216023 0
0.45100330532247 0
0.460023371428918 0
0.469043437535366 0
0.478063503641813 0
0.487083569748261 0
0.496103635854709 0
0.505123701961156 0
0.514143768067604 0
0.523163834174052 0
0.532183900280499 0
0.541203966386947 0
0.550224032493394 0
0.559244098599842 0
0.56826416470629 0
0.577284230812737 0
0.586304296919185 0
0.595324363025633 0
0.604344429132081 0
0.613364495238528 0
0.622384561344976 0
0.631404627451423 0
0.640424693557871 0
0.649444759664319 0
0.658464825770766 0
0.667484891877214 0
0.676504957983662 0
0.68552502409011 0
0.694545090196557 0
0.703565156303005 0
0.712585222409452 0
0.7216052885159 0
0.730625354622348 0
0.739645420728795 0
0.748665486835243 0
0.757685552941691 0
0.766705619048138 0
0.775725685154586 0
0.784745751261034 0
0.793765817367481 0
0.802785883473929 0
0.811805949580377 0
0.820826015686824 0
0.829846081793272 0
0.83886614789972 0
0.847886214006167 0
0.856906280112615 0
0.865926346219063 0
0.87494641232551 0
0.883966478431958 0
0.892986544538406 0
0.902006610644853 0
0.911026676751301 0
0.920046742857749 0
0.929066808964196 0
0.938086875070644 0
0.947106941177092 0
0.956127007283539 0
0.965147073389987 0
0.974167139496435 0
0.983187205602882 0
0.99220727170933 0
1.00122733781578 0
1.01024740392223 0
1.01926747002867 0
1.02828753613512 0
1.03730760224157 0
1.04632766834802 0
1.05534773445446 0
1.06436780056091 0
1.07338786666736 0
1.08240793277381 0
1.09142799888025 0
1.1004480649867 0
1.10946813109315 0
1.1184881971996 0
1.12750826330604 0
1.13652832941249 0
1.14554839551894 0
1.15456846162539 0.0123456790123457
1.16358852773184 0
1.17260859383828 0
1.18162865994473 0
1.19064872605118 0
1.19966879215763 0
1.20868885826407 0
1.21770892437052 0
1.22672899047697 0
1.23574905658342 0
1.24476912268986 0
1.25378918879631 0
1.26280925490276 0
1.27182932100921 0
1.28084938711566 0
1.2898694532221 0
1.29888951932855 0
1.307909585435 0
1.31692965154145 0
1.32594971764789 0
1.33496978375434 0
1.34398984986079 0
1.35300991596724 0
1.36202998207368 0
1.37105004818013 0
1.38007011428658 0
1.38909018039303 0
1.39811024649948 0
1.40713031260592 0
1.41615037871237 0
1.42517044481882 0
1.43419051092527 0
1.44321057703171 0
1.45223064313816 0
1.46125070924461 0
1.47027077535106 0
1.4792908414575 0
1.48831090756395 0
1.4973309736704 0
1.50635103977685 0
1.51537110588329 0
1.52439117198974 0
1.53341123809619 0
1.54243130420264 0
1.55145137030909 0
1.56047143641553 0
1.56949150252198 0
1.57851156862843 0
1.58753163473488 0
1.59655170084132 0
1.60557176694777 0
1.61459183305422 0
1.62361189916067 0
1.63263196526711 0
1.64165203137356 0
1.65067209748001 0
1.65969216358646 0
1.6687122296929 0
1.67773229579935 0
1.6867523619058 0
1.69577242801225 0
1.7047924941187 0
1.71381256022514 0
1.72283262633159 0
1.73185269243804 0
1.74087275854449 0
1.74989282465093 0
1.75891289075738 0
1.76793295686383 0
1.77695302297028 0
1.78597308907672 0
1.79499315518317 0.0123456790123457
};
\addplot [thick, red, opacity=0.4]
table {%
0.194594901979023 0.166666666666667
0.194891783918327 0
0.195188665857631 0
0.195485547796935 0
0.195782429736239 0
0.196079311675543 0
0.196376193614847 0
0.196673075554151 0
0.196969957493455 0.166666666666667
0.19726683943276 0
0.197563721372064 0
0.197860603311368 0
0.198157485250672 0
0.198454367189976 0
0.19875124912928 0
0.199048131068584 0
0.199345013007888 0
0.199641894947192 0
0.199938776886496 0
0.2002356588258 0
0.200532540765104 0
0.200829422704409 0
0.201126304643713 0
0.201423186583017 0.166666666666667
0.201720068522321 0
0.202016950461625 0
0.202313832400929 0.166666666666651
0.202610714340233 0
0.202907596279537 0
0.203204478218841 0.166666666666667
0.203501360158145 0.5
0.203798242097449 0.166666666666651
0.204095124036754 0.166666666666667
0.204392005976058 0
0.204688887915362 0
0.204985769854666 0.166666666666667
0.20528265179397 0.166666666666651
0.205579533733274 0.333333333333333
0.205876415672578 0.333333333333333
0.206173297611882 0.333333333333333
0.206470179551186 0.166666666666667
0.20676706149049 0.499999999999953
0.207063943429794 0.333333333333333
0.207360825369098 0.166666666666667
0.207657707308402 0.5
0.207954589247707 0.166666666666667
0.208251471187011 0
0.208548353126315 0
0.208845235065619 0
0.209142117004923 0.166666666666667
0.209438998944227 0.166666666666667
0.209735880883531 0
0.210032762822835 0.5
0.210329644762139 0.333333333333333
0.210626526701443 0
0.210923408640747 1
0.211220290580051 0.333333333333302
0.211517172519356 0
0.21181405445866 0.833333333333333
0.212110936397964 0.5
0.212407818337268 0.666666666666667
0.212704700276572 0.499999999999953
0.213001582215876 0.333333333333333
0.21329846415518 0.5
0.213595346094484 0.666666666666667
0.213892228033788 0.5
0.214189109973092 0.833333333333255
0.214485991912396 0.666666666666667
0.214782873851701 0.333333333333333
0.215079755791005 0.166666666666667
0.215376637730309 0.166666666666667
0.215673519669613 0.666666666666604
0.215970401608917 0.166666666666667
0.216267283548221 0.333333333333333
0.216564165487525 0.333333333333333
0.216861047426829 0
0.217157929366133 0
0.217454811305437 0.166666666666667
0.217751693244741 0
0.218048575184045 0
0.218345457123349 0.166666666666651
0.218642339062654 0.166666666666667
0.218939221001958 0
0.219236102941262 0.333333333333333
0.219532984880566 0
0.21982986681987 0
0.220126748759174 0
0.220423630698478 0
0.220720512637782 0
0.221017394577086 0
0.22131427651639 0
0.221611158455694 0
0.221908040394999 0
0.222204922334303 0
0.222501804273607 0
0.222798686212911 0
0.223095568152215 0
0.223392450091519 0
0.223689332030823 0
0.223986213970127 0
0.224283095909431 0
0.224579977848735 0
0.224876859788039 0
0.225173741727343 0
0.225470623666647 0
0.225767505605952 0
0.226064387545256 0
0.22636126948456 0
0.226658151423864 0
0.226955033363168 0
0.227251915302472 0
0.227548797241776 0
0.22784567918108 0
0.228142561120384 0
0.228439443059688 0
0.228736324998992 0
0.229033206938297 0
0.229330088877601 0
0.229626970816905 0
0.229923852756209 0
0.230220734695513 0
0.230517616634817 0
0.230814498574121 0
0.231111380513425 0
0.231408262452729 0
0.231705144392033 0
0.232002026331337 0
0.232298908270641 0
0.232595790209946 0
0.23289267214925 0
0.233189554088554 0
0.233486436027858 0
0.233783317967162 0
0.234080199906466 0
0.23437708184577 0
0.234673963785074 0
0.234970845724378 0
0.235267727663682 0
0.235564609602986 0
0.23586149154229 0
0.236158373481595 0
0.236455255420899 0
0.236752137360203 0
0.237049019299507 0
0.237345901238811 0
0.237642783178115 0.166666666666667
0.237939665117419 0
0.238236547056723 0
0.238533428996027 0
0.238830310935331 0
0.239127192874635 0
0.239424074813939 0
0.239720956753244 0
0.240017838692548 0
0.240314720631852 0
0.240611602571156 0
0.24090848451046 0
0.241205366449764 0
0.241502248389068 0
0.241799130328372 0
0.242096012267676 0
0.24239289420698 0
0.242689776146284 0
0.242986658085588 0
0.243283540024893 0
0.243580421964197 0
0.243877303903501 0
0.244174185842805 0
0.244471067782109 0
0.244767949721413 0
0.245064831660717 0
0.245361713600021 0
0.245658595539325 0
0.245955477478629 0
0.246252359417933 0
0.246549241357237 0
0.246846123296542 0
0.247143005235846 0
0.24743988717515 0
0.247736769114454 0
0.248033651053758 0
0.248330532993062 0
0.248627414932366 0
0.24892429687167 0
0.249221178810974 0
0.249518060750278 0
0.249814942689582 0
0.250111824628886 0
0.250408706568191 0
0.250705588507495 0
0.251002470446799 0
0.251299352386103 0
0.251596234325407 0
0.251893116264711 0
0.252189998204015 0
0.252486880143319 0
0.252783762082623 0
0.253080644021927 0
0.253377525961231 0
0.253674407900535 0.166666666666651
};
\node[right, red, rotate=90] at (axis cs:0.27,0.5) {\footnotesize Nominal};
\node[right,black, rotate=90] at (axis cs:0.06,0.5) {{\footnotesize Stochastic}};
\end{axis}

\end{tikzpicture}
\end{center}
\caption{Histogram of the terminal normalized tumor sizes under nominal and stochastic MPC controllers.} \label{histoT}
\end{figure}
\begin{figure}
\begin{center}
\input{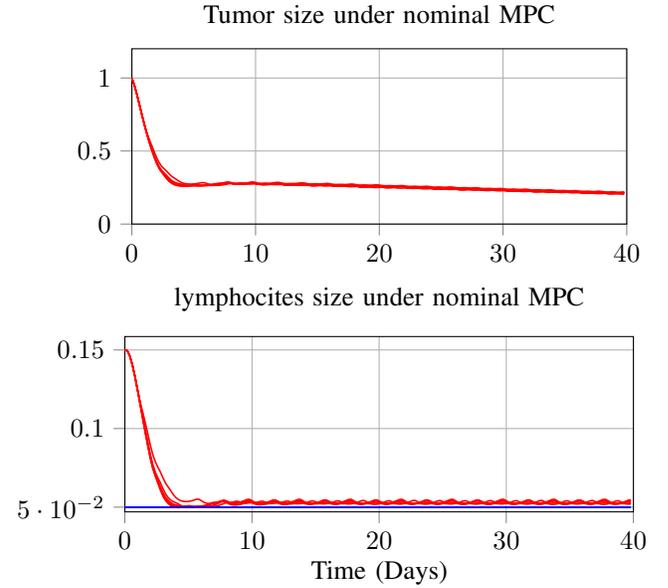}
\end{center}
\caption{Typical closed-loop behavior under the nominal MPC.} \label{behaviorNominal}
\end{figure}
\begin{figure}
\begin{center}
\input{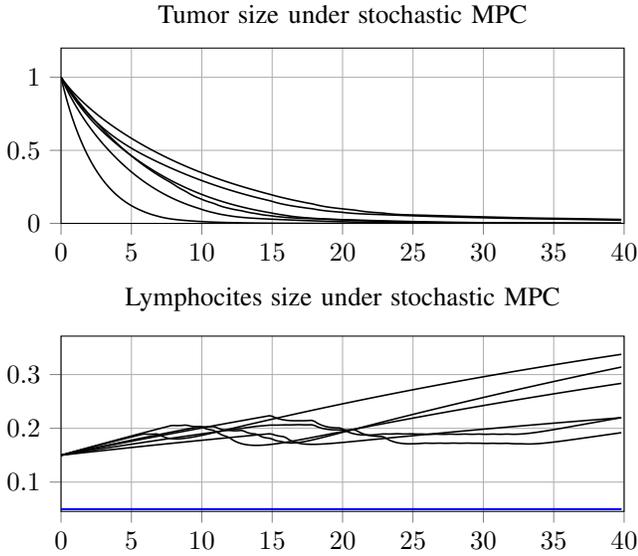}
\end{center}
\caption{Typical closed-loop behavior under the stochastic MPC.} \label{behaviorStochastic}
\end{figure}

Figure \ref{histoT} shows the normalized (w.r.t the maximum bins) histograms of the tumor sizes at the end of the closed-loop simulations. This figure shows that the SNMPC outperforms the nominal MPC as it leads to a vanishing tumor size except for two single outliers where the tumor is increased as explained later on. \\ 
\begin{figure}
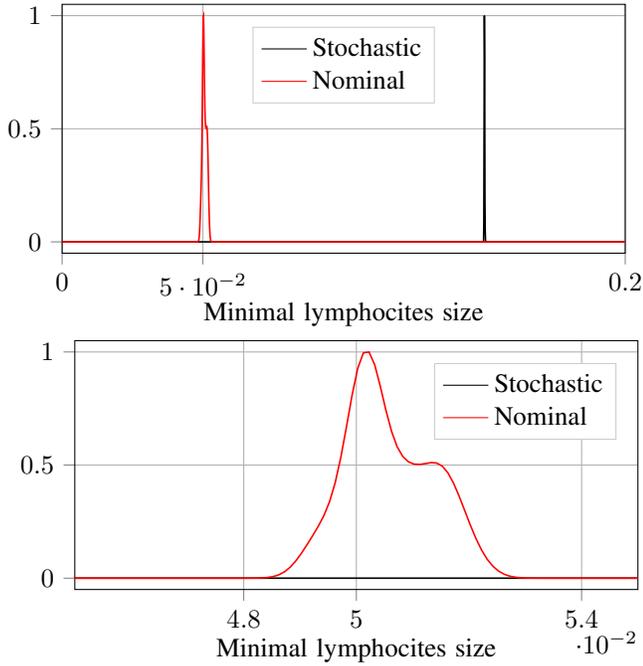

\begin{center}
\input{histoConstr.tex}
\input{histoConstrZoom.tex}
\end{center}
\caption{(top): Histogram of the minimal lymphocytes population size. (Bottom): Zoom on the nominal histogram showing constraints violation in $15\%$ of the scenarios. Note that the bottom plot is a zoom on the top plot around the lower bound $x_2^{min}=0.05$.} \label{histolympho}
\end{figure}
\ \\
Figure \ref{histolympho} shows the normalized histogram of the minimal lymphocytes population's size during the closed-loop simulations. Note how the bottom plot of this figures shows that under the nominal MPC, the constraints is violated in around $15\%$ of the scenarios. Note however how the constraints is largely respected when the SNMPC is used due to the cautious behavior of the stochastic controller. 
\\ \ \\ 
One might notice here that something uncommon is happening as the stochastic controller wins on both sides, namely the cost function and the constraints satisfaction. This can be explained by examining the typical behavior of the closed-loop under the nominal vs the stochastic MPC controllers which are depicted respectively on Figures \ref{behaviorNominal} and \ref{behaviorStochastic}. As a matter of fact, since the nominal controller is not cautious and does not see the risk of violating the constraints, it applies intensive chemotherapy drug from the beginning as this reduced the tumor size quickly and hence lead to a lower value of the cost function. But when the horizon recedes, the closed-loop system is trapped since there is no more possibility to reduce the cost significantly without violating the constraint on the lymphocytes population size. That is the reason the nominal controller can only regulate the lymphocytes size by applying in parallel chemotherapy and immunotherapy (see the Figure \ref{TheUnominal}).
\\ \ \\  

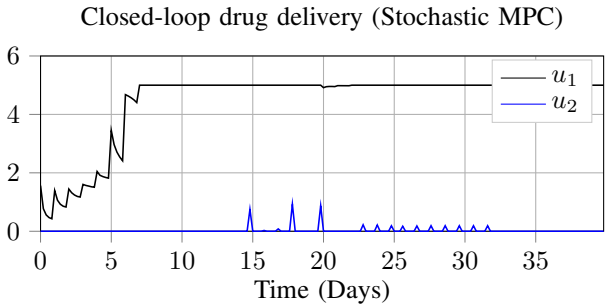
\begin{figure}[H]
\begin{center}
\begin{tikzpicture}

\begin{axis}[
title={Closed-loop drug delivery (Stochastic MPC)},
xlabel={Time (Days)},
xmin=0, xmax=39.8,
ymin=0, ymax=6,
tick align=outside,
tick pos=left,
xmajorgrids,
x grid style={white!69.01960784313725!black},
ymajorgrids,
y grid style={white!69.01960784313725!black},
legend entries={{$u_1$},{$u_2$}},
legend style={draw=white!80.0!black},
legend cell align={left},
width=0.5\textwidth,
height=0.16\textheight
]
\addlegendimage{no markers, black}
\addlegendimage{no markers, blue}
\addplot [semithick, black]
table {%
0 1.56329555498243
0.2 0.77117058518815
0.4 0.558988964775977
0.6 0.468860181669824
0.8 0.427886579498122
1 1.39163770946492
1.2 1.06116370124107
1.4 0.924486056385051
1.6 0.857675449654484
1.8 0.830505737485509
2 1.44481495587162
2.2 1.31850456620897
2.4 1.24113429972566
2.6 1.19259432734254
2.8 1.1714093214957
3 1.60350714561965
3.2 1.56935676917107
3.4 1.54722656122309
3.6 1.52220556929919
3.8 1.50768736492948
4 2.04586565505129
4.2 1.91619213310402
4.4 1.87278631189243
4.6 1.84204479684569
4.8 1.81830097334923
5 3.45002742002964
5.2 2.96518636850649
5.4 2.70931329577187
5.6 2.54091784371895
5.8 2.41385052100762
6 4.67606062031513
6.2 4.62841906345114
6.4 4.57034396773109
6.6 4.49802075818426
6.8 4.40537779707403
7 5
7.2 5
7.4 5
7.6 5
7.8 5
8 5
8.2 5
8.4 5
8.6 5
8.8 5
9 5
9.2 5
9.4 5
9.6 5
9.8 5
10 5
10.2 5
10.4 5
10.6 5
10.8 4.99999999922128
11 5
11.2 5
11.4 5
11.6 4.99999999961615
11.8 4.99999999010732
12 5
12.2 5
12.4 4.9999999986867
12.6 4.99999999048428
12.8 4.99999997908727
13 5
13.2 4.99999999662978
13.4 4.99999998922386
13.6 4.9999999793706
13.8 4.99999996561664
14 4.99999999333292
14.2 4.9999999863053
14.4 4.99999997723362
14.6 4.99999996507417
14.8 4.9999999483155
15 4.99999998305896
15.2 4.99999997590331
15.4 4.99999996653137
15.6 4.99999995389711
15.8 4.99999993613955
16 4.9999999774773
16.2 4.99999996917211
16.4 4.99999995842417
16.6 4.99999994404961
16.8 4.99999992398026
17 5
17.2 5
17.4 5
17.6 5
17.8 4.99999999786927
18 4.99999995814098
18.2 4.99999994899235
18.4 4.99999993695159
18.6 4.99999992071918
18.8 4.99999989798918
19 4.99999995557337
19.2 4.99999994506201
19.4 4.99999993146944
19.6 4.9999999133376
19.8 4.99999988879439
20 4.91562576827797
20.2 4.94818264438863
20.4 4.95633182063207
20.6 4.95631658521706
20.8 4.95202226723622
21 4.98361552917015
21.2 4.98309977304451
21.4 4.98249486829833
21.6 4.98181462054499
21.8 4.98107360125459
22 4.9999999435615
22.2 4.99999993158411
22.4 4.9999999163015
22.6 4.99999989618522
22.8 4.9999998687788
23 4.99999993521465
23.2 4.99999992253886
23.4 4.99999990631379
23.6 4.99999988490959
23.8 4.999999855699
24 4.99999992755051
24.2 4.99999991413276
24.4 4.9999998969417
24.6 4.99999987425644
24.8 4.99999984328448
25 4.99999991991894
25.2 4.9999999056158
25.4 4.99999988728857
25.6 4.99999986325321
25.8 4.99999983098793
26 4.99999991260693
26.2 4.99999989737092
26.4 4.99999987787271
26.6 4.99999985233197
26.8 4.99999981811088
27 4.99999990442201
27.2 4.99999988824067
27.4 4.99999986753709
27.6 4.99999984042811
27.8 4.99999980414598
28 4.99999989572977
28.2 4.99999987858748
28.4 4.99999985665297
28.6 4.99999982793111
28.8 4.99999978949021
29 4.99999988665966
29.2 4.99999986851281
29.4 4.99999984529232
29.6 4.99999981488549
29.8 4.99999977418604
30 4.99999987717577
30.2 4.9999998579743
30.4 4.99999983340551
30.6 4.99999980123707
30.8 4.99999975817406
31 4.99999986616804
31.2 4.99999984556115
31.4 4.99999981970315
31.6 4.99999978785457
31.8 4.99999974275106
32 4.99999986082168
32.2 4.99999984003253
32.4 4.99999981349832
32.6 4.99999977862309
32.8 4.99999973089201
33 4.99999985437414
33.2 4.99999983231178
33.4 4.99999980427789
33.6 4.99999976753411
33.8 4.99999971734671
34 4.99999984440797
34.2 4.99999982096668
34.4 4.99999979123503
34.6 4.99999975231751
34.8 4.99999969921104
35 4.99999983155814
35.2 4.99999980653702
35.4 4.99999977482566
35.6 4.99999973333966
35.8 4.99999967675065
36 4.99999981583906
36.2 4.99999978896509
36.4 4.99999975491634
36.6 4.99999971038289
36.8 4.99999964964716
37 4.99999979699503
37.2 4.99999976793349
37.4 4.99999973111859
37.6 4.99999968297267
37.8 4.99999961731581
38 4.99999977469913
38.2 4.99999974306626
38.4 4.99999970299521
38.6 4.99999965059161
38.8 4.99999957912864
39 4.99999974830866
39.2 4.99999971363764
39.4 4.99999966971898
39.6 4.99999961228453
39.8 4.99999953396185
};
\addplot [semithick, blue]
table {%
0 0
0.2 0
0.4 0
0.6 0
0.8 0
1 0
1.2 0
1.4 0
1.6 0
1.8 0
2 0
2.2 0
2.4 0
2.6 0
2.8 0
3 0
3.2 0
3.4 0
3.6 0
3.8 0
4 0
4.2 0
4.4 0
4.6 0
4.8 0
5 0
5.2 0
5.4 0
5.6 0
5.8 0
6 0
6.2 0
6.4 0
6.6 0
6.8 0
7 0
7.2 0
7.4 0
7.6 3.50977973990534e-09
7.8 1.27893794985578e-08
8 0
8.2 0
8.4 0
8.6 4.4176359785532e-09
8.8 1.81948609221752e-08
9 0
9.2 0
9.4 0
9.6 4.65718748014977e-09
9.8 2.2034726995137e-08
10 0
10.2 0
10.4 0
10.6 6.64102151404153e-09
10.8 2.83760024779043e-08
11 0
11.2 3.11879199004034e-10
11.4 0
11.6 1.29027914845194e-08
11.8 5.53007849860642e-08
12 0
12.2 1.93621749273988e-09
12.4 0
12.6 1.32415512975369e-08
12.8 6.91558076513776e-08
13 0
13.2 4.96901457497406e-09
13.4 0
13.6 1.99844437649797e-08
13.8 1.63720573594717e-07
14 0
14.2 0
14.4 1.58542461876246e-08
14.6 4.32810940159715e-08
14.8 0.750973584552405
15 0
15.2 0
15.4 4.7675084624846e-09
15.6 2.58686610571947e-08
15.8 0.0272678904143903
16 0
16.2 3.94061595884965e-10
16.4 8.24543137056561e-09
16.6 3.52332446922287e-08
16.8 0.0806419603976113
17 0
17.2 0
17.4 0
17.6 1.1470896783849e-08
17.8 0.926180733445156
18 6.04869791065985e-10
18.2 6.42339554671856e-09
18.4 1.98479604278113e-08
18.6 6.54447558395703e-08
18.8 6.44359994275478e-07
19 3.11680761621494e-09
19.2 1.08293996344896e-08
19.4 2.98915258743351e-08
19.6 1.06343071337321e-07
19.8 0.872165464252275
20 0
20.2 0
20.4 0
20.6 0
20.8 0
21 0
21.2 0
21.4 0
21.6 0
21.8 0
22 1.21837168766763e-08
22.2 2.77997536997264e-08
22.4 7.31977734245149e-08
22.6 3.47798813650554e-07
22.8 0.221101556144392
23 1.54679311264717e-08
23.2 3.48787703116826e-08
23.4 9.64046337203797e-08
23.6 6.21426737075495e-07
23.8 0.213769156374337
24 1.97601519593696e-08
24.2 4.47109856899686e-08
24.4 1.34478976000178e-07
24.6 2.28026726467173e-06
24.8 0.187223948927195
25 2.44560845749575e-08
25.2 5.57740056210499e-08
25.4 1.81229537115459e-07
25.6 0.171160664442639
25.8 1.6782694367066e-06
26 2.93292984970078e-08
26.2 6.71971563706859e-08
26.4 2.28868163552593e-07
26.6 0.1817654534812
26.8 6.83838402531532e-07
27 3.50611819420043e-08
27.2 8.15765390751492e-08
27.4 2.98709585609157e-07
27.6 0.189828070244727
27.8 4.40951181095e-07
28 4.20883805825977e-08
28.2 1.0068300102015e-07
28.4 4.13561207795398e-07
28.6 0.190050648415788
28.8 3.30454508546832e-07
29 5.08788098703236e-08
29.2 1.27176126575112e-07
29.4 6.35938795400332e-07
29.6 0.189475599382507
29.8 2.67543137319876e-07
30 6.20823966031975e-08
30.2 1.65838620938816e-07
30.4 1.2476662019274e-06
30.6 0.189385885902683
30.8 2.26953725372856e-07
31 6.49243594469276e-08
31.2 1.23198428933094e-07
31.4 1.14507787053976e-05
31.6 0.189406865294862
31.8 1.91614279749089e-07
32 7.56447843858654e-08
32.2 1.93781048129494e-07
32.4 8.44977109120546e-07
32.6 6.95617779303604e-07
32.8 1.43583548823059e-07
33 3.54570862366858e-08
33.2 5.60053114460087e-08
33.4 7.87389012356118e-08
33.6 8.02595774902082e-08
33.8 5.47482860316103e-08
34 1.94510618992172e-08
34.2 2.71620209444804e-08
34.4 3.40925627082622e-08
34.6 3.55145435128043e-08
34.8 2.90292817029152e-08
35 1.14768201601025e-08
35.2 1.54461426559504e-08
35.4 1.88246919413399e-08
35.6 1.9908194479322e-08
35.8 1.74782682982467e-08
36 6.64351763026082e-09
36.2 9.08409859341494e-09
36.4 1.12418107297634e-08
36.6 1.22023886911401e-08
36.8 1.10648467156818e-08
37 1.85507342886328e-09
37.2 3.72354333355678e-09
37.4 6.06739907271827e-09
37.6 7.77056661875144e-09
37.8 6.97353299358654e-09
38 2.13247740605825e-09
38.2 3.28532493539433e-09
38.4 4.22251553699638e-09
38.6 4.68623057871049e-09
38.8 4.44554779870175e-09
39 5.52450581486274e-10
39.2 1.52419324607458e-09
39.4 2.27179337912384e-09
39.6 2.65770839118866e-09
39.8 2.58189854034729e-09
};
\end{axis}

\end{tikzpicture}
\end{center}
\caption{Typical drug delivery under stochastic MPC.} \label{TheUstochastic}
\end{figure}

\ \\
On the other hand, the stochastic MPC does not fall in this {\em trap} as handling all the clusters representative makes it aware of a high risk of constraints violation in case intensive chemotherapy is used from the beginning. That is the reason why, it applies chemotherapy only after a while when the level of lymphocytes becomes high enough to ensure a secure delivery of chemotherapy drug. This can be clearly seen on Figures \ref{behaviorStochastic} and \ref{TheUstochastic}. 
\\ \ \\ 
Note that a longer prediction horizon could have brought the nominal controller into the same strategy than the stochastic one avoiding thus the above mentioned trap. The choice of the scenario is here to illustrate the difference between the two settings and the capabilities of SNMPC to enforce the satisfaction of the constraints when compared to a nominal MPC. \\ \ \\ 
Finally, table \ref{tablecpu} shows the comparison between the statistics of the computation time that is needed to solve the underlying optimization problems at each control updating period. This table clearly shows that using $n_{cl}=3$ cluster induces on average an extra computational burden of $40\%$ while increasing the dispersion of the computation to a higher extent. This also suggest that without using the clustering approach, the computation time would be too prohibitive. 

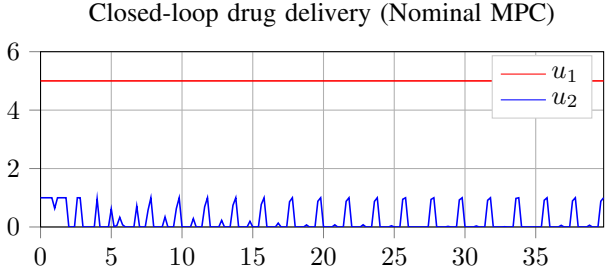
\begin{figure}
\begin{center}
\begin{tikzpicture}

\begin{axis}[
title={Closed-loop drug delivery (Nominal MPC)},
xmin=0, xmax=39.8,
ymin=0, ymax=6,
tick align=outside,
tick pos=left,
xmajorgrids,
x grid style={white!69.01960784313725!black},
ymajorgrids,
y grid style={white!69.01960784313725!black},
legend entries={{$u_1$},{$u_2$}},
legend style={draw=white!80.0!black},
legend cell align={left},
width=0.5\textwidth,
height=0.16\textheight
]
\addlegendimage{no markers, red}
\addlegendimage{no markers, blue}
\addplot [semithick, red]
table {%
0 5
0.2 5
0.4 5
0.6 5
0.8 5
1 5
1.2 5
1.4 5
1.6 5
1.8 4.99999999453066
2 5
2.2 5
2.4 5
2.6 5
2.8 4.99999999933432
3 5
3.2 5
3.4 5
3.6 5
3.8 5
4 5
4.2 5
4.4 5
4.6 5
4.8 5
5 5
5.2 5
5.4 5
5.6 5
5.8 5
6 5
6.2 5
6.4 5
6.6 5
6.8 5
7 5
7.2 5
7.4 5
7.6 5
7.8 5
8 5
8.2 5
8.4 5
8.6 5
8.8 5
9 5
9.2 5
9.4 5
9.6 5
9.8 5
10 5
10.2 5
10.4 5
10.6 5
10.8 5
11 5
11.2 5
11.4 5
11.6 5
11.8 5
12 5
12.2 5
12.4 5
12.6 5
12.8 5
13 5
13.2 5
13.4 5
13.6 5
13.8 5
14 5
14.2 5
14.4 5
14.6 5
14.8 5
15 5
15.2 5
15.4 5
15.6 5
15.8 5
16 5
16.2 5
16.4 5
16.6 5
16.8 5
17 5
17.2 5
17.4 5
17.6 5
17.8 4.9999999999686
18 5
18.2 5
18.4 5
18.6 5
18.8 5
19 5
19.2 5
19.4 5
19.6 5
19.8 4.99999999904062
20 5
20.2 5
20.4 5
20.6 5
20.8 5
21 5
21.2 5
21.4 5
21.6 5
21.8 4.99999999806432
22 5
22.2 5
22.4 5
22.6 5
22.8 4.99999999945187
23 5
23.2 5
23.4 5
23.6 5
23.8 4.99999999704595
24 5
24.2 5
24.4 5
24.6 5
24.8 5
25 5
25.2 5
25.4 5
25.6 5
25.8 4.99999999591992
26 5
26.2 5
26.4 5
26.6 5
26.8 4.99999999752865
27 5
27.2 5
27.4 5
27.6 5
27.8 4.99999999484498
28 5
28.2 5
28.4 5
28.6 5
28.8 4.99999999642797
29 5
29.2 5
29.4 5
29.6 5
29.8 4.99999999372902
30 5
30.2 5
30.4 5
30.6 5
30.8 4.99999999530483
31 5
31.2 5
31.4 5
31.6 5
31.8 4.99999999257676
32 5
32.2 5
32.4 5
32.6 5
32.8 4.99999999676226
33 5
33.2 5
33.4 5
33.6 4.99999999963227
33.8 4.99999999131317
34 5
34.2 5
34.4 5
34.6 5
34.8 4.99999999310019
35 5
35.2 5
35.4 5
35.6 4.99999999869727
35.8 4.99999999016247
36 5
36.2 5
36.4 5
36.6 5
36.8 4.99999999176843
37 5
37.2 5
37.4 5
37.6 4.9999999976164
37.8 4.99999998890445
38 5
38.2 5
38.4 5
38.6 5
38.8 4.99999999180664
39 5
39.2 5
39.4 5
39.6 4.99999999896374
39.8 4.99999999047993
};
\addplot [semithick, blue]
table {%
0 1
0.2 1
0.4 1
0.6 1
0.8 1
1 0.636884788793795
1.2 0.999999997339128
1.4 1
1.6 1
1.8 1
2 8.28239896022621e-09
2.2 2.68668713784047e-08
2.4 1.17516569922765e-07
2.6 0.999999713923546
2.8 0.999999846787163
3 2.44403136669854e-08
3.2 3.85525961014565e-08
3.4 6.36540727480841e-08
3.6 1.20924419357721e-07
3.8 3.33719677854559e-07
4 0.961909134714885
4.2 1.1101469653037e-07
4.4 6.79567284901966e-08
4.6 7.21762972156136e-08
4.8 1.79528981915592e-07
5 0.614179122495177
5.2 2.23652154929851e-06
5.4 0.0552100746399763
5.6 0.328677101349773
5.8 0.0736340842163154
6 2.16023260130083e-07
6.2 2.06705016205181e-07
6.4 2.72186170211282e-07
6.6 5.62344889563137e-07
6.8 0.712635239964616
7 1.08675030572957e-07
7.2 1.74514085413002e-07
7.4 4.07132582183078e-07
7.6 0.576636392527282
7.8 0.999998997449115
8 9.62114981034821e-08
8.2 1.36666980836094e-07
8.4 2.24494800702133e-07
8.6 5.25026525185224e-07
8.8 0.344290838982349
9 7.84803633398973e-08
9.2 1.32299282062169e-07
9.4 3.20407108141905e-07
9.6 0.633320376988044
9.8 0.999999183142151
10 7.32022111099305e-08
10.2 1.07959394588368e-07
10.4 1.83246793569323e-07
10.6 4.39483524501525e-07
10.8 0.285731213974595
11 6.03982504936429e-08
11.2 1.04846941443919e-07
11.4 2.59779435575646e-07
11.6 0.692261397437781
11.8 0.999999327559113
12 6.39320159038298e-08
12.2 9.60826842210274e-08
12.4 1.6571876581567e-07
12.6 4.02160042882185e-07
12.8 0.22654869043654
13 5.97687647528938e-08
13.2 1.04432116530761e-07
13.4 2.60337770530183e-07
13.6 0.740791848385412
13.8 0.999999316966732
14 5.86077108840903e-08
14.2 8.91701981279038e-08
14.4 1.55434783611074e-07
14.6 3.80274487392426e-07
14.8 0.193594824122582
15 5.64627225790279e-08
15.2 9.94629417501678e-08
15.4 2.49554400640785e-07
15.6 0.769467984915911
15.8 0.999999341219877
16 4.39764044317199e-08
16.2 6.80997468792624e-08
16.4 1.20631042543608e-07
16.6 2.98827092011072e-07
16.8 0.133138596983441
17 5.31898574458236e-08
17.2 9.44309777320766e-08
17.4 2.38642949476846e-07
17.6 0.864620317510606
17.8 0.999999362150112
18 5.66660200979451e-08
18.2 8.7171435442809e-08
18.4 1.53085523406232e-07
18.6 3.75715630778532e-07
18.8 0.0705365660109888
19 5.29617868320147e-08
19.2 9.39831575164821e-08
19.4 2.37499276151153e-07
19.6 0.880426121471084
19.8 0.999999365411689
20 5.52718251710749e-08
20.2 8.52923492266545e-08
20.4 1.50276343972321e-07
20.6 3.69963814369756e-07
20.8 0.0736668881277227
21 5.27526064163066e-08
21.2 9.38141453950133e-08
21.4 2.37402795706479e-07
21.6 0.878007018569191
21.8 0.999999365329427
22 5.44985329897431e-08
22.2 8.42806434344759e-08
22.4 1.48836057656839e-07
22.6 3.67223196865859e-07
22.8 0.0757771976372406
23 5.28536708101057e-08
23.2 9.41173649209492e-08
23.4 2.3836032441196e-07
23.6 0.876097944958198
23.8 0.999999362604567
24 4.33562422487528e-08
24.2 6.77536381527504e-08
24.4 1.20853019967351e-07
24.6 3.00626766929362e-07
24.8 0.0413107820309562
25 5.18336697681388e-08
25.2 9.26031560512e-08
25.4 2.35303292879278e-07
25.6 0.948127007903091
25.8 0.999999365869308
26 4.24748478998691e-08
26.2 6.18453948264896e-08
26.4 9.76463092497642e-08
26.6 1.82065208060465e-07
26.8 5.05636610464321e-07
27 5.29129139455382e-08
27.2 9.42495231412657e-08
27.4 2.38947787008528e-07
27.6 0.942834952487133
27.8 0.999999357944516
28 5.56238871485104e-08
28.2 8.62353850181969e-08
28.4 1.52553438631106e-07
28.6 3.76493667374406e-07
28.8 0.0207195894329838
29 5.35664128792676e-08
29.2 9.54629600238752e-08
29.4 2.42019017823271e-07
29.6 0.925941786994129
29.8 0.999999350428204
30 5.54065564539261e-08
30.2 8.59528527014218e-08
30.4 1.52237380783623e-07
30.6 3.76375583970668e-07
30.8 0.0346577649835821
31 5.42751457768113e-08
31.2 9.67440197334103e-08
31.4 2.45222171729348e-07
31.6 0.913313291838322
31.8 0.999999342473003
32 4.48221158906014e-08
32.2 7.01027119737924e-08
32.4 1.25144284376789e-07
32.6 3.11161078235359e-07
32.8 0.00926391139558255
33 5.37118687231667e-08
33.2 9.59620598236291e-08
33.4 2.43830813150856e-07
33.6 0.977269665115067
33.8 0.999999342033422
34 4.373818479804e-08
34.2 6.37820871164832e-08
34.4 1.00899251323152e-07
34.6 1.88512520137148e-07
34.8 5.24134769320432e-07
35 5.60004978039051e-08
35.2 9.95729852340384e-08
35.4 2.5190462149332e-07
35.6 0.914497661124033
35.8 0.999999325898535
36 5.57412976876657e-08
36.2 8.65391977272314e-08
36.4 1.53579393667947e-07
36.6 3.80747943978868e-07
36.8 0.0567458462213679
37 5.6473671026515e-08
37.2 1.00651941045409e-07
37.4 2.5494556165509e-07
37.6 0.893626739036391
37.8 0.999999317642735
38 5.05248667726156e-08
38.2 7.87206974434864e-08
38.4 1.40240341508318e-07
38.6 3.4890982959162e-07
38.8 0.0610544884290446
39 4.95923639826337e-08
39.2 8.92583364427741e-08
39.4 2.27501241604239e-07
39.6 0.881138722207328
39.8 0.999999394927345
};
\end{axis}

\end{tikzpicture}
\end{center}
\caption{Typical drug delivery under nominal MPC.} \label{TheUnominal}
\end{figure}

\begin{table}
\begin{center}
\begin{tabular}{|l|c|c|}
  \hline
   & {\bf Mean} (ms) & {\bf Standard deviation} (ms)\\
  \hline
  {\bf Nominal} & 270 & 76 \\
  {\bf Stochastic} & 383 & 330 \\
  \hline
\end{tabular}
\end{center}
\caption{Statistics of the computation times (in ms) of a single solution of the associated optimization problem. (CasADi \citep{Andersson2018} (python version) on a MacBookPro 2.9 GHz Intel Core i7).} \label{tablecpu}
\end{table}

\section{Conclusion and future work} \label{secConc}
In this paper, a clustering-based framework is proposed to derive an approximated version of nonlinear stochastic MPC control design that is illustrated on a realistic and challenging examples involving a high dimensional non reconstructible uncertainty vector. Work in progress targets a better understanding of way the number of clusters can be rationally chosen, the impact of the choice of the labels involved in the clustering step, the size of the FIFO buffer (the forgetting rate of previous samples). Application to many other real-life examples is also under investigation. 
\\ \ \\ 
\bibliography{stochasticMPC.bib}
\bibliographystyle{aaai-named}

\end{document}